\documentclass[conference]{IEEEtran}
\IEEEoverridecommandlockouts
\usepackage{cite}
\usepackage{amsmath,amssymb,amsfonts}
\usepackage{algorithmic}
\usepackage{graphicx}
\usepackage{textcomp}
\usepackage{url}
\usepackage{xcolor}
\usepackage{listings}
\usepackage{fancyvrb}
\usepackage{tikz}

\usepackage{array}
\newcolumntype{L}[1]{>{\raggedright\let\newline\\\arraybackslash\hspace{0pt}}m{#1}}
\newcolumntype{C}[1]{>{\centering\let\newline\\\arraybackslash\hspace{0pt}}m{#1}}
\newcolumntype{R}[1]{>{\raggedleft\let\newline\\\arraybackslash\hspace{0pt}}m{#1}}

\def\BibTeX{{\rm B\kern-.05em{\sc i\kern-.025em b}\kern-.08em
    T\kern-.1667em\lower.7ex\hbox{E}\kern-.125emX}}
\begin{document}

\title{Conceptualizing A Configuration Service for Complex Automation Systems}

\author{
\IEEEauthorblockN{{\bf Javad Ghofrani}\\ \tt\footnotesize javad.ghofrani@htw-dresden.de} \and
\IEEEauthorblockN{{\bf Paul Patolla}\\ \tt\footnotesize paul.patolla@htw-dresden.de} \and 
\IEEEauthorblockN{{\bf Daniel Richter}\\ \tt\footnotesize daniel.richter@htw-dresden.de} \and
\IEEEauthorblockN{{\bf Dirk Reichelt}\\ \tt\footnotesize dirk.reichelt@htw-dresden.de} \and 
}

\maketitle

\begin{abstract}
Arrowhead Framework (AHF) is being developed to enable large-scale IoT based automation by providing an interoperability layer for local clouds. This framework aims to  create an abstract model for distributed, heterogeneous, and non-linear systems. Managing the variability in such environments plays a key role in handling complex automation tasks such as in smart production systems. However, there is no standard solution available for handling the variability and configuration specifications in such environments. In this paper, we analyze the existing solutions for configuration management in industrial automation frameworks and provide leverage points for a standardization framework for handling configurations of automated production systems based on the concept of industrial internet of things. 
 
 
\end{abstract}

\section{Introduction}\label{sec:introduction}
\IEEEPARstart{A}{utomation} of industrial systems is one of the main goals of the Industry 4.0 concept which is growing with a fast pace\cite{lasi2014industry,russmann2015industry}. Based on the concept of Industry 4.0, more automation will be provided by better connectivity technology. Sensor systems and IoT devices make it possible to perform more monitoring and controlling over production processes and plants which leads to major increases in productivity and quality. However, handling the complexity of such connectivity and automation requires additional tools and methods. Arrowhead Framework (AHF) is an established tool to handle the connectivity and provide automation in industrial systems. It  provides an abstraction layer over the distributed, heterogeneous and fully autonomous automation systems. Through this abstraction, which is based on Service Oriented Architecture, various systems and sub systems can communicate and collaborate. AHF provides discovery and orchestration services for finding and consuming services within and between systems~\cite{delsing2017iot}.Through a ServiceRegistry module, provided by AHF, a service consumer can search for certain services. Information about service providers---metadata---are stored in the service registry when a service provider enters the Arrowhead cloud. Service consumers can search and find possible service providers using search criteria on this metadata. Service consumers should decide which service to consume according to their requirements. However, this kind of service consumption offers not much flexibility. It is fixed how the service provider can work and provide some services. Within this mechanism, it is not foreseen how the consumer and provider can vary to satisfy the required conditions.  In this paper we address the problem of the flexibility in IoT automation framework Arrowhead. in this regard, we specify the requirements for applying flexibility and variability in providing and consuming services in the context of industrial Internet of Things. As first step we investigate the existing mechanisms and tools for providing updates and changes in IoT devices connected to the automation clouds. They promise to handle the reconfiguration of systems in local clouds and certain domains. Most of these software put focus on rolling out some updates using agent based mechanisms with considering the security and safe installation of updates on IoT devices. However, these technologies and tools have their limits. Since these tools are developed for managing the updates, they could be triggered by administrators for a range of devices or systems when the functionality of services or devices in IoT environment should be updated.
Enabling the flexibility for service provider and service consumer leads to a more optimized usage of resources and increases the productivity of IoT enabled systems. However, handling the changes on devices in a manual way is time consuming, complicated and error prone due to the complexity and permanent changing environment like IoT automation management clouds.  Our contribution is to propose  an approach for applying automation into the update / change management of IoT automation systems. To this end, service providers and service consumers in AHF should be able to anticipate changes and reconfigure their functionalities without human interference.  

\section{Configuration in IoT: Existing Approaches}\label{sec:existing}
Carlsson et al.~\cite{carlsson2016configuration} address the need for a configuration server to deal with  variability in the heterogeneous environments. In their proposed specification, a configuration service as a part of IoT automation platform should distribute updates to the IoT devices. based on their specifications, the configuration server should be able to handle three variability level on IoT devices and services, i.e., distributing updates of firmware, operating system, and programs. There are various solutions which are developed to cover various aspects of variability and configurability in such systems~\cite{abbas2018multi,chatzigiannakis2012true}. Using software product line engineering (SPLE)\cite{weiss1999software} methods is an established method to handle variability in software systems. Ayala et al.~\cite{ayala2015software} address the variability in  multi agent based solutions in IoT. They utilize SPL methods to support reusablity of assets in the development of agent developed for IoT purposes. Using multi agent based solutions in IoT system is a common approach~\cite{ayala2016using,di2018iot,liu2017distributed}. However, the majority of these studies consider the heterogeneity of environment and try to find a solution to adapt the IoT devices to the changes in environment. Most of the papers in this domain assume that the environment is changing and the agents are constant part that should anticipate changes and adapt themselves.  In this regard, existing studies neglected the variability and configuration between the cooperating agents in the industrial IoT context. It is clear that due to the huge connectivity and heterogeneity in an IoT based environment, especially in systems of systems, there is no guarantee that the agents will be developed by a central development unit. Therefore, handling the variability in such way that proposed in aforementioned studies is not possible. 

The challenge of the configuration of IoT devices in industrial automation systems has been approached by three major projects in the Eclipse IoT space. However, each project has a different focus and thus varying strategies for the propagation of configuration to the devices are being employed. In the following sections an analysis of Eclipse Kapua~\cite{eclipseKapua}, Eclipse Kura~\cite{eclipseKura}, and Eclipse hawkBit~\cite{hawkbit} is being done, in order to expose possible paths for future development. For simplicity's sake, the prefix 'Eclipse' is going to be omitted in future sections of this paper.

Kura is a middleware focussed on building IoT gateways and transmitting acquired data to IoT cloud platforms via MQTT. It enables remote management of such gateways by offering a set of REST APIs. Configuration of devices with Kura is based on an extensible framework utilizing the Java framework Spring. In order to access device configurations a generic device agent is extended with a set of plugins. These plugins expose device features to the platform, e.g. polling rate, camera enabled etc. While this approach allows for the modular design of more complex IoT applications, mere configuration and update tasks are being bloated. As IoT devices are constrained by limited storage and networking capabilities, the utilization of Java Plugins is not a solution that is lightweight enough for industrial automation scenarios.

The Kapua project encompasses the broadest scope of the three as its focus lies on the comprehensive management of IoT devices and sensors. It covers the complete lifecycle from connectivity to configuration and operations. Kapua's Device Management component enables remote operations on connected devices. An open, stack-ignorant contract based on MQTT is being utilized as an interface to the devices. Kura can be used to implement this kind of communication protocol allowing for the management and provisioning of device configurations, applications, and services. When looking at large scale industrial automation scenarios however, scalability and automatability of configuration rollouts is not something neither Kapua alone, nor in cooperation with Kura, can provide in the context of configurations and updates.

hawkBit aims at solving the issue of scalability in the domain of software updates allowing for large scale orchestration of update campaigns. Single update artifacts can be bundled together into software modules. A bundle of software modules is called a distribution, which is the primary entity being rolled out to a defined number devices by hawkBit. The update server itself implements it's own content delivery network for the provisioning of content. hawkBit uses a polling based mechanism for the transmission of updates, i.e. the hawkBit client on a device pulls matching software updates periodically. While the employment of a matching ledger between update packages and groups of devices is a step forward in achieving a more scalable architecture, highly automated systems cannot rely on hawkBit for re-configuration. This is because a human operator is still required to create and manage the update campaigns.


\begin{figure*}
\centering
\includegraphics[width=\textwidth]{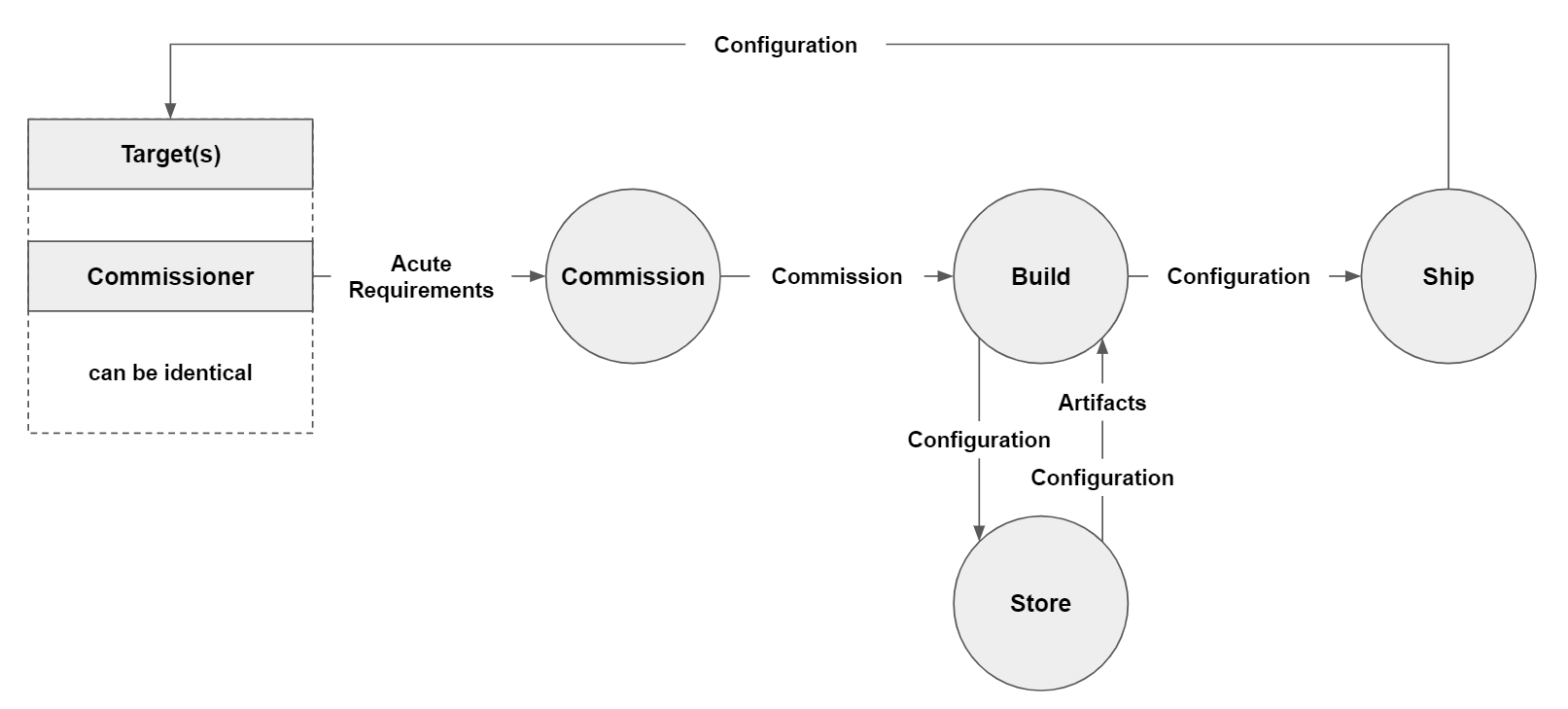}
\caption{Configuration Process Data Flow}
\label{fig_flow}
\end{figure*}

\section{Configuration Process Analysis}\label{sec:specifications}
AHF enables service consumers to look for service providers providing the services that match their requirements. In a fully automated environment, service consumers and service providers operate autonomously and with minimal human interaction. If requirements cannot be fulfilled by the current system configuration status, service consumers should therefore be able to reconfigure their service providers accordingly. This has several implications on the architecture of an automation cloud. The following sections provide a view into the most important parts of the configuration process and deduces specifications for an automated system, which can fulfill these process steps. The data flow of proposed configuration system can be examined in Figure \ref{fig_flow}.

\subsection{Commissioning Configuration}\label{sec:requirements}
System (re)-configuration begins with the definition of requirements, which can happen explicitly or implicitly. Requirements for system configurations originate from two sources. Firstly, they can be generated by a system component itself should it detect some change in critical parameters. More often however, requirements for system configuration change because of some outside influence, which may be changes in process definitions or business rules as defined by organizational management positions. The reason for the prevalence of this classical approach of mainly information system external requirements definition may in large part be due the lack of a fitting approach for system self configuration. Although configuration requirements should be stated explicitly, a major advantage of an automated configuration system lies in providing a method for interpreting implicitly stated requirements automatically. Nevertheless, a configuration system  needs to have interfaces to both, internal and external sources of requirements in order to be able to produce a matching system configuration. 

IoT devices fulfill tasks as part of a larger process chain, which is set in place to create some kind of added value or fulfill some higher goal, i.e. in industrial environments the smart production of goods. Related tasks fulfilling their own sub-goals can be grouped together into corresponding business scenarios. In a fully automated environment, the configuration of a number of related IoT devices then can be viewed as the representation of such business scenario. The matching business scenario should thusly be referenced in a configuration rollout plan as defined by requirements. This way, dependencies between devices corresponding to a certain business scenario can be modelled. The definition of a business scenario contains boundary constraints which need to be fulfilled at all times so that the scenario stays intact. These constraints ensure that no change in configuration of a device belonging to the scenario threatens the stability of the overarching process chain. In the context of AHF-compliant systems, knowledge about business scenarios could be managed by the PlantDescription system.

To minimize problems created by conflicting requirements, the requirements themselves need a measure of importance attached to them, so that the configuration system can then satisfy them according to policy. One such way to measure importance could be a simple attachment of a positive integer value. The requirement with the highest attached value for a target device or scenario then overrides all others. To prevent arising inflexibilities due to a strictly static approach to configuration scheduling, more measures have to be attached to a requirement. A requirement then becomes more than just the description of a specific configuration, but also contains additional descriptive and instructive metadata. Another way of scheduling configuration rollouts could be introducing an economical model building on the concept of autonomy of service providers and consumers inside a local automation cloud. Using such model, system participants issuing requirements get an importance currency, which is being renewed periodically in an amount matching their organisational or processual gravity. They can then use this importance currency to determine the relative importance of their commissioned configuration. Priority then becomes a good, which market participants bid for. It can be argued, whether this approach solves the issue of determining the most important configurations for a specific device, because it only produces sound results, if the available currency is distributed exactly to the real importance of the participating parties. Determining this importance is still a hard task to accomplish and could probably only be solved by finding a statistical optimum using simulation methods. The first proposed method is a special case of the market-based approach with unlimited currency per participant. A mapping of importance to configuration contents and targets has to be determined in each organisation separately. Should this not be possible, heuristic scheduling methods such as first-in-first-out can be used as an emergency measure. They should be used with care however, as they are ignorant to the semantics of the requirements and the actual configuration contents. A way of dealing with possibly harmful configurations is by defining a set of constraints for a business scenario, which need to be complied with, in order to secure system and process health. As long as these constraints are not being infringed, it does not matter which configuration is being run first, or at all. Following definition can be derived from the preceding analysis: A requirement is an abstract commission of configuration. Each commission can require multiple configurations, but for each configuration there is only one source commission. Dimensions of such commission are:
\newtheorem{dimension}{Dimension}
\begin{dimension}[Importance]
How urgent is the configuration? As discussed above, this value can be determined implicitly or explicitly.
\end{dimension}
\begin{dimension}[Time]
When should the configuration take place and when should it revert? Not every configuration needs to be executed immediately, but can be scheduled to a later time on purpose. This might be the case, if some larger processual change is also taking place at a specific point in time or when there is some recurring pattern in reconfiguration. Reverting configurations can especially be used when the required change in service is only planned for a short period of time.
\end{dimension}
\begin{dimension}[Target]
Which singular devices or business scenarios should be configured?
\end{dimension}
\begin{dimension}[Type]
Which services should targets be able to provide after configuration and on which level? A service can be a sensing or an acting task. As the service quality is also being described, security requirements be represented accordingly.
\end{dimension}

The description of a configuration commission, i.e. a requirement, does not include more technical details than needed, because to the potential service consumer they are irrelevant. It is merely an abstract description of a required service, which could also be a security requirement on a specific level or the availability of a specific operating system version. The relationships between abstract requirements and technical implications is being handled by the core of the configuration system proposed in this paper: the configuration factory.

\subsection{Building Configuration}\label{sec:build}
At the heart of the proposed configuration system specifications lies a component which interprets the requirements provided by the requirements interface and tries to fulfill them. One of the main obstacles in enabling fully autonomous systems in Industry 4.0 is the still widespread need for human involvement. Previous sections have shown, that human operators are still relevant in creating and managing configurations using current state-of-the-art solutions. \cite{eclipseKapua, eclipseKura, hawkbit} Therefore, an approach is being proposed that minimizes said human interaction.

The configurability of a system depends largely on the quality of its architecture. Configuration assets should be reused based on explicitly stated and managed variation points. \cite{vanderLinden2007spl} Model Driven Architecture (MDA) comes into play, because it also builds on the foundations of well described abstracted system components: models. Following the principles of MDA, abstract capability models could serve as the basis for the automatic generation of configuration based on the requirements. Before configuration for target devices can be built then, a meta-model of the organisation's device inventory needs to be created, which describes potentially provided services and other features such as firmware and communication or storage capabilities for each possible device on a high level. \cite{schmidt2002poa} As stated earlier, some invariants -- factors that may not be changed -- have to be defined as well to ensure the proper functioning of the device or business scenario. Lowering the level of abstraction, the AHF DeviceRegistry could be used as the representation of the current system configuration status yielding information about possible pre-conditions, which may need to be met. Using the requirements as descriptions of commissioned post-conditions, a contract can then be given a generator component, which builds a configuration file matching the target devices and operating system. \cite{davidfrankel2003} This can only be done, if a relationship between each desired configuration outcome, device and operating system is being maintained. In the language of MDA, these relationships are called transformation components. \cite{omg2014mdaguide} These transformation components bridge the gap between the abstract definitions of requirements and the device specific configuration artifact, which can be in varying formats. In summary, following models need to be created on descending levels of abstraction to enable an automatic generation of configuration content:

\subsubsection{Device Feature Metamodels (DFM)}\label{sec:modelmeta}
This model lies on the highest level of abstraction. It contains the overarching structure of an IoT device's general feature space and can be reused across organisations as a template for deriving an organisational device feature model. A device on the level of this metamodel can be defined as a 7-tuple $D = (C, M, T, P, \Sigma, A, \Omega)$, where:
\begin{description}[\IEEEsetlabelwidth{$C, M, T$}\IEEEusemathlabelsep]
    \item [$C, M, T$] Are computational, memory and communication capabilities as defined in \cite{delsing2017iot}.
    \item [$P$] Power constraints, which can be specified by needed power supply, self-sufficiency etc.
    \item [$\Sigma, A$] Sensing and acting capabilities, respectively. These describe any features enabling the device to interact with its environment apart from communication.
    \item [$\Omega$] An operating system, which is the platform the device operates on. This has implications on the way the device handles messages -- thus also configuration instructions -- internally.
\end{description}

\subsubsection{Organisational Feature Models (OFM)}\label{sec:modeldevice}
Models for specific IoT devices from the various manufacturers, which are used inside an organisation specify the schema for describing devices. These models extend the structure of the meta-model in a way that possible variation points and in which form they may be accessed are being defined. In the XML-inspired example of a specific device description below, it can be examined that for a certain organisation the number of cores and clock speed might be of interest for configuration while the basic structure of the DFM prevails.\newline

\begin{Verbatim}[frame=single,fontsize=\small{},label=EXAMPLE DEVICE DESCRIPTION DERIVED FROM OFM]
<Device>
    <ID>RPI3_B_ARM_01</ID>
    <Capabilities>
        <Computational>
            <Cores>4</Cores>
            <Clock>1.2 GHz</Clock>
        </Computational>
    </Capabilities>
    ...
</Device>
\end{Verbatim}

\subsubsection{Device State Models}\label{sec:modelstate}
The actual states of the devices' aforementioned variation points need to be maintained in a ledger, in order to match these states (pre-conditions) with possible requirements (post-conditions). As these device states match the real-time conditions of the devices inside the local cloud this representation lies on the lowest level of abstraction of the three discussed models. In the case of AHF, the DeviceRegistry system could provide a means for storing device states as part of device metadata. However, this metadata has to be updated periodically, in order to retain the necessary level of currentness.

In summary, the DFM proposed in \ref{sec:modelmeta} should be used as a blueprint to build an organisation-specific OFM extending the basic structure to suit configuration needs. The OFM can then be used further to describe single devices, which serve as the foundation to capture device variability, actual states and  thus possible targets for configuration. Table \ref{table_factory_data} gives an overview of required services and data objects for building configuration files based on the models described in this section. The actual product of the configuration build process is still to be specifically determined. As the target devices diverge in platforms, so does the form of configuration instructions range from shell scripts to JSON-files. One important aspect of the built configuration package is that it should contain metadata about its shipping requirements. These are for instance required battery charge, system interrupts, criticality, latest shipping time etc. 

\begin{table}
\renewcommand{\arraystretch}{2.5}
\caption{Relevant Input Data For Configuration Factory}
\label{table_factory_data}
\centering
\begin{tabular}{p{2.5cm}|p{2.5cm}|p{2.5cm}}
\hline
\bfseries Source Service & \bfseries Input Data & \bfseries Relevant Content\\
\hline\hline
RequirementsInterface & Configuration Commission & Service requirements and constraints\\
\hline
DeviceRegistry & Device State & Target device status and platform\\
\hline
PlantDescription & Business Scenario & Scenario constraints for target device(s)\\
\hline
ArtifactStore & Transformation Component & Requirements translation schema\\
\hline

\end{tabular}
\end{table}

\subsection{Storing Configuration}\label{sec:store}
Not only device states need to be readily accessible, but as the proposed configuration system uses various models to construct configuration, it might prove beneficial to processing load to store both, configuration artifacts used in constructing the configuration executables, as well as past built executable configuration instructions. The benefits for the latter are twofold: Firstly, newly built configurations might become the golden standard for future reconfiguration as described by \cite{delsing2017iot}. Secondly, snapshots of system configuration to revert to add an additional layer of security should new configuration be to the detriment of system performance or overall stability. The more obvious reason however for using a storage mechanism in an MDA-oriented system, is for providing a persistent location to the build artifacts, most prominently transformation components, which map required capabilities to targeted device platforms. The Configuration system in AHF provides a ConfigurationStore service to devices, in order to retrieve assigned configuration instructions. This approach however raises two problems. On one hand, pull-only configuration is not a sufficient mechanism for important and interdependently modelled system changes. On the other hand, the schema of an implementation of such storage service needs to accommodate not only for finalized configuration executables, but also for the various artifacts proposed in this paper. Following the principle of separation of concerns, a separate storage and retrieval service for these kinds of artifacts might prove useful, e.g. an ArtifactStore service which provides the configuration factory with both, the target device schema and subsequently required transformation components.

\subsection{Shipping Configuration}\label{sec:ship} 
The process of shipping configuration can benefit the most from the already established configuration tools for IoT devices discussed in \ref{sec:existing}. Shipping configuration can be broken down into two consecutive aspects. The first step in configuration shipment comprises the physical transmission of executables from the device hosting the configuration service to configurable target devices. This is followed by a process of configuration execution on the target devices and thereby the final satisfaction of the original requirements starting the configuration process chain.

The transmission of configuration files is subject to two major restrictions: Security and network traffic limitations of both, the devices themselves as well as the organisational infrastructure capabilities as a whole. Methods for securing the transmission and the contents of configuration files depend largely on the used protocols, but Transport Layer Security (TLS) can be used for a multitude of communication protocols. It is however not feasible for resource-constrained devices. For authorization and authentication only, the standard solutions offered by the AMQP and MQTT protocols can be used. 

While security only targets one quality of transmission, the question of a propagation strategy in the local cloud still stands. It cannot be answered so easily as it builds on a number of key assumptions about the architecture of the configuration service itself. As part of this paper's specifications, it is assumed that configurable devices are technically capable of running an interface of the configuration service as an agent, while not limiting their main sensing and acting capabilities. Although this approach discriminates against a number of resource-constrained devices, it enables a larger degree of flexibility in possible strategy choice. For one, a one-to-many transmission between a configuration broker and target devices is being enabled by such approach. This would mirror the approach hawkBit uses. The challenge of network constraints for large scale configuration rollouts could be solved using an approach similar to the BitTorrent protocol, where a select number of available nodes seed files across the entire network. The cost of rolling out configuration would be distributed across the devices inside the local cloud as opposed to an approach where a single server distributes the files to target devices creating large network spikes. The configuration system still has to determine which devices are computationally able and have sufficient power resources available, in order to perform seeding.

An agent-based approach enables a standardized interpretation of configuration metadata in accord with the real-time device state. The successful execution of the configuration instruction should be ended with a final notice to the requirements interface, so the configuration request can then be marked as completed.

\section{Conclusion and Future Work}\label{sec:conclusion}
Concluding, the specification of a configuration service is as complex a task as the systems that require it. This paper presented an analysis of the most important steps in configuration for IoT devices and proposed key approaches and methods in handling the complexity in automation systems. Future work could be directed towards deriving specific artifacts and specifications from the presented abstract models for an organisational use case. Especially the creation of transformation artifacts for a selection of possible configuration options and well-established target platforms would be step forward in enabling the MDA-based methods. An evaluation can then be done, whether the proposed approaches are actually feasible in production. A further study of the edge field between SPLE, MDA and requirements engineering could provide useful insights for the creation of an actual configuration tool, e.g. \cite{ReinhartzBerger2019}.

\bibliographystyle{IEEEtran}
\begingroup
  \small 
  \bibliography{references}
\endgroup

\end{document}